\documentclass[11pt]{article}

\usepackage{arxiv}

\usepackage[utf8]{inputenc}
\usepackage[T1]{fontenc}
\usepackage{url}
\usepackage{booktabs}
\usepackage{longtable}
\usepackage{fvextra}
\usepackage{amsmath}
\usepackage{amsfonts}
\usepackage{graphicx}
\usepackage{float}
\usepackage{microtype}
\usepackage[numbers,sort&compress]{natbib}
\usepackage{xcolor}
\usepackage{hyperref}
\usepackage{doi}
\usepackage{cleveref}
\usepackage{fontawesome5}

\graphicspath{{data/figures/}}

\title{PrefBench: Evaluating Zero-Shot LLM Agents in Hidden-Preference Personalized Pricing Negotiations}

\author{
  Yingjie Lei\\
  University of Aberdeen\\
  \texttt{u19yl22@abdn.ac.uk}
}

\date{May 2026}

\renewcommand{\shorttitle}{PrefBench}

\hypersetup{
  pdftitle={PrefBench: Evaluating Zero-Shot LLM Agents in Hidden-Preference Personalized Pricing Negotiations},
  pdfsubject={personalized pricing, negotiation, LLM agents, benchmark, hidden preferences},
  pdfauthor={Yingjie Lei},
  pdfkeywords={personalized pricing, LLM agents, agent evaluation, hidden buyer preferences, simulator-based benchmark}
}

\begin{document}
\maketitle

\begin{center}
\small \faGithub\ Code: \url{https://github.com/ChaosTheProducer/PrefBench}
\end{center}

\begin{abstract}
Personalized pricing negotiations are a challenging testbed for LLM agents because successful interaction does not guarantee profitable decision making. A seller may produce valid actions and close many deals while still pricing poorly when buyer willingness to pay and bargaining traits remain hidden. This paper presents \textbf{PrefBench}, a simulator-based benchmark for hidden-preference personalized pricing negotiations. Each episode pairs a simulated buyer with a fixed vehicle-customization bundle; the seller observes public persona descriptors, bundle information, and negotiation history, while latent buyer variables govern valuation, patience, counter-offer behavior, and walkaway decisions. PrefBench evaluates this setting through an LLM-facing state-summary protocol that constrains agents to return strict JSON actions under a fixed hidden-information boundary. We evaluate zero-shot LLM sellers against heuristic references over 7,500 episodes. The tested LLMs follow the protocol reliably and achieve deal rates above 0.99, but their seller-profit outcomes remain weak: the best LLM average profit is only slightly above the random baseline and far below a simple concession heuristic under the same episode stream. These results show that structured action compliance and agreement-seeking behavior can coexist with weak profit-sensitive bargaining. PrefBench provides a controlled benchmark for evaluating pricing-agent behavior under hidden buyer preferences.

\end{abstract}

\keywords{Personalized pricing \and LLM agents \and Agent evaluation \and Hidden buyer preferences \and Simulator-based benchmark}

\section{Introduction}
\label{sec:introduction}

Personalized pricing is difficult when buyers value the same product differently and the seller observes only a partial view of those differences. In configurable-product negotiations, willingness to pay can depend on budget, use case, feature priorities, aesthetic taste, and bargaining style. The seller must therefore decide how aggressively to price without observing the buyer's latent valuation.

This setting creates a useful distinction for evaluating seller agents. A high offer may preserve margin but trigger rejection or walkaway; a low offer may close the deal quickly but sacrifice profit. The central question is not only whether an agent can reach agreement, but whether it can use a short negotiation to turn partial information into profitable agreement.

This distinction is especially relevant for LLM agents. In pricing negotiations, valid actions and completed deals are not sufficient evidence of good decision making: a seller agent can obey a JSON schema and close nearly every deal while still pricing too softly. Hidden-preference pricing therefore separates structured action compliance from profit-sensitive strategy.

Existing benchmarks do not directly isolate this combination. Dynamic and personalized pricing work studies demand response, revenue optimization, and customer heterogeneity \citep{choi2023SemiParametricContextualPricing,liu2021DynamicPricingEcommerce,biggs2021ModelDistillationRevenue,dube2017PersonalizedPricingConsumer}. Retail simulators provide controlled environments for pricing or recommendation policies \citep{xia2023RetailSynthSyntheticData}. Negotiation benchmarks study bargaining behavior, preferences, and mental-state reasoning \citep{baarslag2012FirstAutomatedNegotiating,lin2014GeniusIntegratedEnvironment,xia2024MeasuringBargainingAbilities,chan2024NegotiationToMBenchmarkStresstesting}, while general LLM-agent benchmarks study tool use, task completion, and rule following \citep{liu2023AgentBenchEvaluatingLLMs,qin2023ToolLLMFacilitatingLarge,yao2024TauBenchToolAgentUser}. What is missing is a fixed evaluation setting that combines hidden buyer variables, seller-side profit, multi-round pricing actions, and structured LLM outputs under one shared protocol.

We present PrefBench, a controlled simulator-based benchmark for hidden-preference personalized pricing negotiations. PrefBench stress-tests seller-side pricing behavior under a fixed hidden-information boundary, using simulator-level outcomes to support reproducible comparison. Each episode pairs one simulated buyer with one fixed vehicle-customization bundle. The agent observes public persona descriptors, bundle information, and negotiation history, while the simulator retains benchmark-defined latent variables such as willingness to pay, price sensitivity, patience, counter strength, walkaway threshold, and feature preferences. Agents act through \texttt{offer}, \texttt{accept}, and \texttt{walkaway} moves, and outcomes are evaluated by simulator seller profit together with negotiation diagnostics.

Using PrefBench, we evaluate zero-shot LLM agents under a strict structured-action protocol. At each decision point, the model receives a prompt rendered from the current observable state summary rather than a persistent free-form negotiation transcript. In our experiments, LLMs produce zero invalid episodes and deal rates above 0.99, showing that the tested models can follow the interface and reach agreements. The profit results tell a different story. The best LLM average profit is 6,749.21 USD, only slightly above the random baseline at 6,572.33 USD and far below a simple concession heuristic at 14,774.11 USD under the same episode stream. These results do not show that LLMs are generally poor at pricing. They show that, in this benchmark, high action compliance and high agreement rate can coexist with weak seller-profit behavior.

The main contributions are:
\begin{itemize}
  \item We formulate a hidden-preference personalized-pricing negotiation task for evaluating LLM agents as seller-side decision makers.
  \item We introduce a controlled benchmark with public-data-informed persona descriptors, benchmark-defined latent buyer variables, a structured product catalog, frozen evaluation splits, and shared episode-level metrics.
  \item We define an LLM-facing protocol that renders only observable state summaries into prompts and constrains agents to return strict JSON actions under the same hidden-information boundary.
  \item We report zero-shot LLM baselines showing that protocol compliance and agreement rate can substantially overstate seller-side pricing performance.
  \item We release PrefBench as a controlled evaluation substrate for pricing-agent research, with simulator assumptions and scope boundaries documented explicitly.
\end{itemize}

\section{Related Work}
\label{sec:related_work}

\paragraph{Dynamic and personalized pricing.}
Dynamic pricing has been studied through contextual pricing, demand estimation, and learning-based price adjustment \citep{choi2023SemiParametricContextualPricing,liu2021DynamicPricingEcommerce,pandey2020DeepReinforcementLearning}. Personalized pricing additionally raises questions about consumer heterogeneity, fairness perception, and how individual or segment-level information should be used \citep{biggs2021ModelDistillationRevenue,dube2017PersonalizedPricingConsumer,priester2020SpecialPriceJust}. Product attributes and configurable options have also been studied as differentiated pricing objects, including automobile variants and configurable products \citep{thomassen2017EmpiricalModelAutomobile,wang2022AssortmentPlanningPricing}. PrefBench builds on this motivation but isolates a different evaluation question: how a seller-side agent prices a fixed configurable bundle when buyer valuation and bargaining traits remain hidden.

\paragraph{Negotiation agents and environments.}
Automated negotiation has a long history in multi-agent systems, including shared platforms and competitions such as ANAC and Genius \citep{baarslag2012FirstAutomatedNegotiating,lin2014GeniusIntegratedEnvironment}. NegMAS provides another research-oriented framework for constructing negotiation simulations \citep{mohammad2021NegMASPlatformAutomated}. Recent LLM-facing negotiation work evaluates bargaining ability and theory-of-mind reasoning in negotiation settings \citep{xia2024MeasuringBargainingAbilities,chan2024NegotiationToMBenchmarkStresstesting}. PrefBench focuses this line on a pricing-specific question: whether an LLM seller can convert observable buyer and interaction signals into seller-profit outcomes when the evaluator retains hidden buyer valuation variables. The unit of evaluation is action-level pricing behavior over configurable product features, rather than negotiation reasoning alone.

\paragraph{LLM agents and structured action evaluation.}
General LLM-agent benchmarks evaluate whether models can act across interactive environments, use tools, follow task rules, or call APIs reliably \citep{liu2023AgentBenchEvaluatingLLMs,qin2023ToolLLMFacilitatingLarge,yao2024TauBenchToolAgentUser}. PrefBench shares this structured-interface concern: LLM agents must convert an observable state-summary prompt into a valid JSON action. It then evaluates whether a model that follows the protocol can also make profit-sensitive pricing decisions when buyer preferences remain hidden.

\paragraph{Simulation benchmarks and persona grounding.}
Standardized simulator environments are useful when real interaction data are costly, sensitive, or difficult to control. Synthetic or simulator-based benchmarks are used in retail and pricing simulation settings \citep{xia2023RetailSynthSyntheticData,villarrubia-martin2025DynamicPricingHighSpeed}. PrefBench follows this simulator-based tradition while making its assumptions explicit. Its observable buyer descriptors are informed by public U.S. demographic, income, and travel-purpose sources \citep{CensusProfileUnited,bureauIncomeUnitedStates,bricka2024summary}, while hidden preference and bargaining variables are benchmark-defined rather than directly observed. This separation positions PrefBench as a controlled evaluation benchmark for hidden-preference pricing behavior.

\section{Task Formulation}
\label{sec:task_formulation}

\begin{figure}[t]
  \centering
  \includegraphics[width=0.98\linewidth]{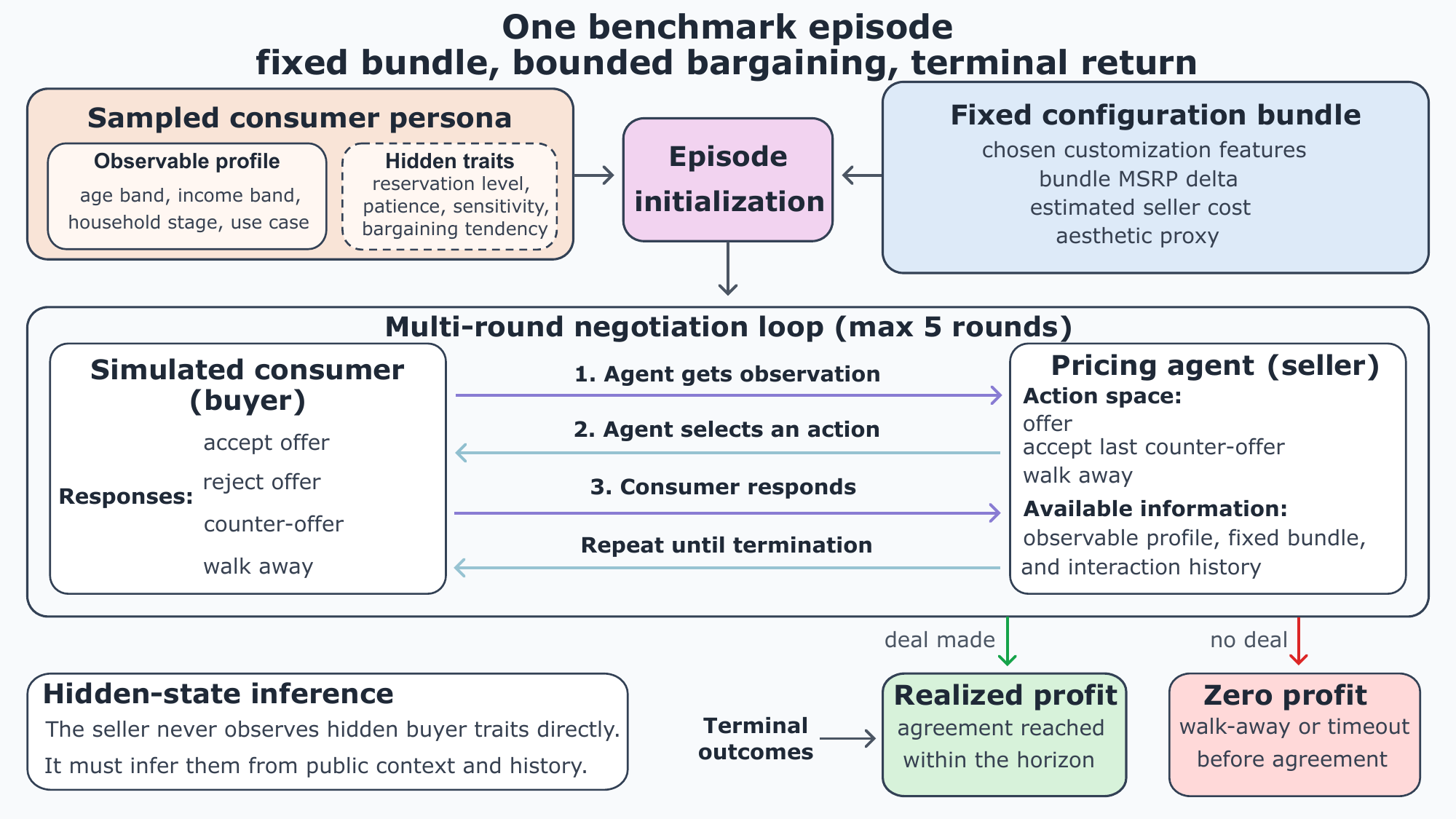}
  \caption{Structure of one PrefBench episode. A sampled buyer persona and a fixed customization bundle define the hidden-preference negotiation problem; the seller observes only public profile, bundle, and history information before choosing pricing actions.}
  \label{fig:episode_overview}
\end{figure}

PrefBench defines a finite-horizon seller-side pricing task over a bounded vehicle-customization scope. As summarized in \cref{fig:episode_overview}, one episode is one negotiation between a pricing agent and a simulated buyer over one fixed customization bundle. The instantiated catalog, described in \cref{sec:simulator_assets} and summarized in \cref{app:customization_scope}, contains 20 canonical options across 11 customization dimensions. Each episode samples one option per dimension to form the bundle under negotiation.

The bundle identity is fixed during the episode, so the agent adjusts price and negotiation moves while observing buyer responses. This task design focuses the benchmark on pricing under hidden preferences, with product recommendation, assortment design, package-dependency logic, and vehicle-configuration optimization left outside the episode.

The seller observes only a partial state. At the start of an episode, the observation contains the selected bundle and coarse buyer profile fields. As the episode unfolds, the observation also includes the current round, the latest seller offer, the latest buyer response, any buyer counter-offer, and a visible history summary. The seller does not observe the buyer's latent willingness to pay, reservation-price level, price sensitivity, patience, counter strength, walkaway threshold, or feature-preference weights. These variables remain on the simulator side and shape the buyer's responses. \Cref{tab:task_specification} summarizes the benchmark-facing information boundary.

\begin{table}[t]
  \centering
  \caption{Compact PrefBench task specification. The seller receives only visible fields, while hidden buyer variables are retained by the simulator.}
  \label{tab:task_specification}
  \small
  \begin{tabular}{p{0.24\linewidth}p{0.68\linewidth}}
    \toprule
    Component & Fields or definitions \\
    \midrule
    Customization scope & Fixed vehicle-customization bundle sampled from 20 canonical options across 11 dimensions; one option is selected per dimension and remains fixed throughout the episode. Package dependencies and cross-option compatibility constraints are not modeled. \\
    Visible buyer fields & Age band, income band, household stage, ownership stage, and primary use case. \\
    Visible bundle fields & Selected option keys and dimensions, bundle MSRP delta, estimated implementation cost, and aesthetic proxy score. \\
    Visible interaction fields & Current round, remaining seller turns, latest seller offer, latest buyer response, latest buyer counter-offer if available, and history length. \\
    Hidden buyer variables & Willingness-to-pay process, reservation-price level, price sensitivity, patience, counter strength, walkaway threshold, and feature-preference weights. \\
    Seller actions & \texttt{offer(price\_offer\_usd)}, \texttt{accept}, and \texttt{walkaway}. \\
    Core reported metrics & Realized seller profit, deal rate, average rounds, and LLM invalid-output rate. \\
    \bottomrule
  \end{tabular}
\end{table}

The action space has three move types: \texttt{offer}, \texttt{accept}, and \texttt{walkaway}. An \texttt{offer} proposes a price for the fixed bundle. The legal offer domain is any non-negative numeric USD value; offers are not rounded to integer dollars, clipped to an MSRP-derived interval, or constrained to exceed implementation cost. This choice keeps price aggressiveness observable in the outcome: below-cost accepted offers are legal but can produce negative deal profit. An \texttt{accept} commits to the buyer's latest counter-offer when such a counter exists. A \texttt{walkaway} ends the negotiation from the seller side. The buyer may accept, reject, counter, or walk away in response to seller offers.

Output validity and action availability are separated. For LLM policies, parser-invalid outputs include malformed JSON, non-object responses, unsupported move types, missing offer prices, non-numeric offer prices, and negative offer prices. These outputs terminate only the current episode and are counted in the reported invalid-output metric. Legal terminal outcomes, including seller walkaway, buyer walkaway, and horizon timeout, are not invalid actions. A schema-valid action can still be unavailable in the current state: for example, \texttt{accept} requires an existing buyer counter-offer. If no counter-offer is available, the simulator records the action as an unsuccessful environment step rather than as a parser-invalid output. In the benchmark used here, the maximum negotiation horizon is \(H=5\) seller decision rounds.

The primary seller objective is realized profit,
\begin{equation}
  \mathrm{profit\_usd} = p_{\mathrm{deal}} - c_{\mathrm{impl}},
\label{eq:profit}
\end{equation}
where \(p_{\mathrm{deal}}\) is the final agreed price and \(c_{\mathrm{impl}}\) is the estimated implementation cost of the selected bundle. This objective makes agreement frequency and economic value distinct: an agent can close many deals by pricing softly, while another may preserve margin at the cost of more failed negotiations.

No-deal outcomes contribute zero realized profit. Deal rate, average rounds, and invalid-output rate are therefore reported as diagnostic metrics rather than as replacements for the seller objective. This distinction is important for interpreting LLM agents: a policy that maximizes immediate agreement can look successful under a completion metric while still performing poorly under seller profit.

The benchmark profit is terminal realized seller profit, not the auxiliary environment reward used internally by the simulator. The environment may assign event-level rewards or penalties to intermediate outcomes such as rejected offers, counter-offers, walkaway, timeout, or unavailable actions. These values are useful for simulator traces and learned-agent training, but the main tables report \(\mathrm{profit\_usd}\) from \cref{eq:profit} for deal outcomes and zero profit for no-deal outcomes.

The task can be viewed as a finite-horizon partially observable Markov decision process (POMDP). The simulator maintains the full latent negotiation state, including buyer valuation and bargaining traits, while the seller receives only partial observations and chooses actions that affect both future feedback and terminal return. PrefBench uses this formulation to define the information boundary; the LLM agents are evaluated as prompt-based policies rather than trained POMDP solvers.

\section{Simulator and Benchmark Assets}
\label{sec:simulator_assets}

PrefBench instantiates each pricing episode with a fixed vehicle-customization catalog. The catalog starts from selected Mercedes-Benz E350 configuration options and MSRP deltas from the official build interface \citep{BuildYourOwn}, then maps them into the benchmark product space summarized in \cref{app:customization_scope}. Each option carries three benchmark attributes: a consumer-facing MSRP delta, a seller-side implementation-cost proxy, and an aesthetic prior. For a selected bundle \(b\), the implementation cost is
\begin{equation}
  c_{\mathrm{impl}}(b) = 0.5 \sum_{i \in b} \Delta \mathrm{MSRP}_i ,
\label{eq:implementation_cost}
\end{equation}
where \(\Delta \mathrm{MSRP}_i\) is the consumer-facing MSRP delta of option \(i\). The 0.5 multiplier defines a fixed seller-side cost proxy, which makes realized seller profit comparable across policies under the same bundle distribution. The bundle is sampled once at the start of an episode and remains fixed, so the agent's decision is price negotiation over a given product configuration.

\begin{figure}[t]
  \centering
  \includegraphics[width=0.98\linewidth]{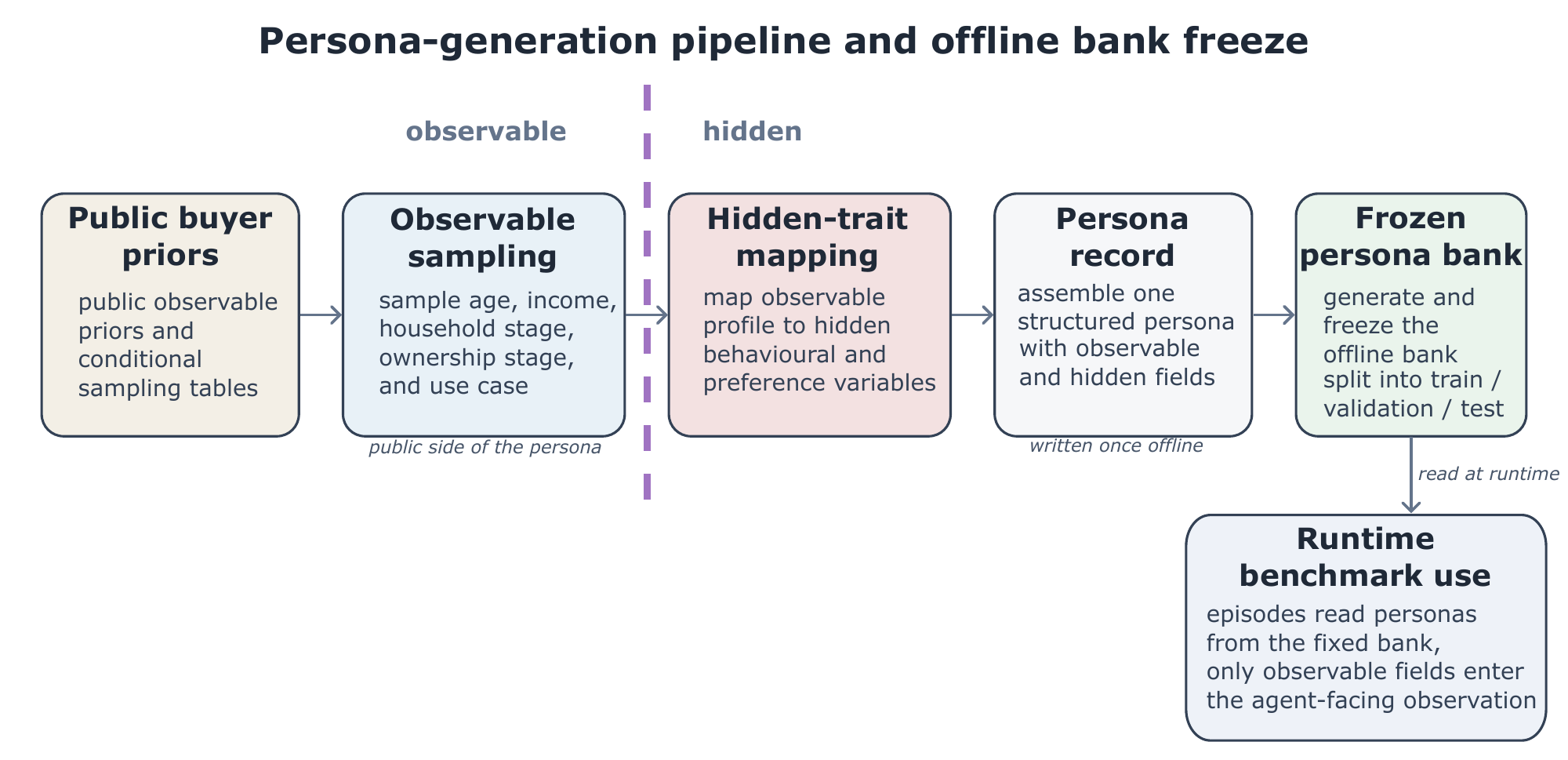}
  \caption{Persona-bank construction. Observable descriptors are sampled from public-data-informed priors, hidden traits are generated by benchmark-defined mappings, and a fixed evaluation split is reused across benchmark runs.}
  \label{fig:persona_pipeline}
\end{figure}

The buyer population is represented by a frozen semi-synthetic persona bank. \Cref{fig:persona_pipeline} summarizes the construction pipeline. Observable fields are coarse profile descriptors, including age band, income band, household stage, ownership stage, and primary use case, with U.S. demographic, income, and travel-purpose sources used as anchors \citep{CensusProfileUnited,bureauIncomeUnitedStates,bricka2024summary}. The observable profile is sampled through conditional tables: income band, household stage, and ownership stage depend on age band, while primary use case depends on household stage. Hidden traits are then generated from the observable profile through benchmark-defined conditional mappings, numeric mixtures, bounded shifts, and coupling rules. These hidden traits create controlled partial observability for the pricing task. The full conditional tables are provided in \cref{app:persona_conditionals}.

Each persona record separates seller-visible descriptors from simulator-internal buyer variables. Hidden variables include reservation value, price sensitivity, aesthetic sensitivity, patience, counter strength, walkaway threshold, brand loyalty, impulsivity, belief obscurity, and feature-preference weights. These variables govern how different buyers value the same fixed bundle while remaining hidden from the seller. The experiments in this paper use a fixed 7,500-record evaluation split generated with seed 123 and stratified by age and income; all methods are evaluated on this same split.

\begin{figure}[t]
  \centering
  \includegraphics[width=0.98\linewidth]{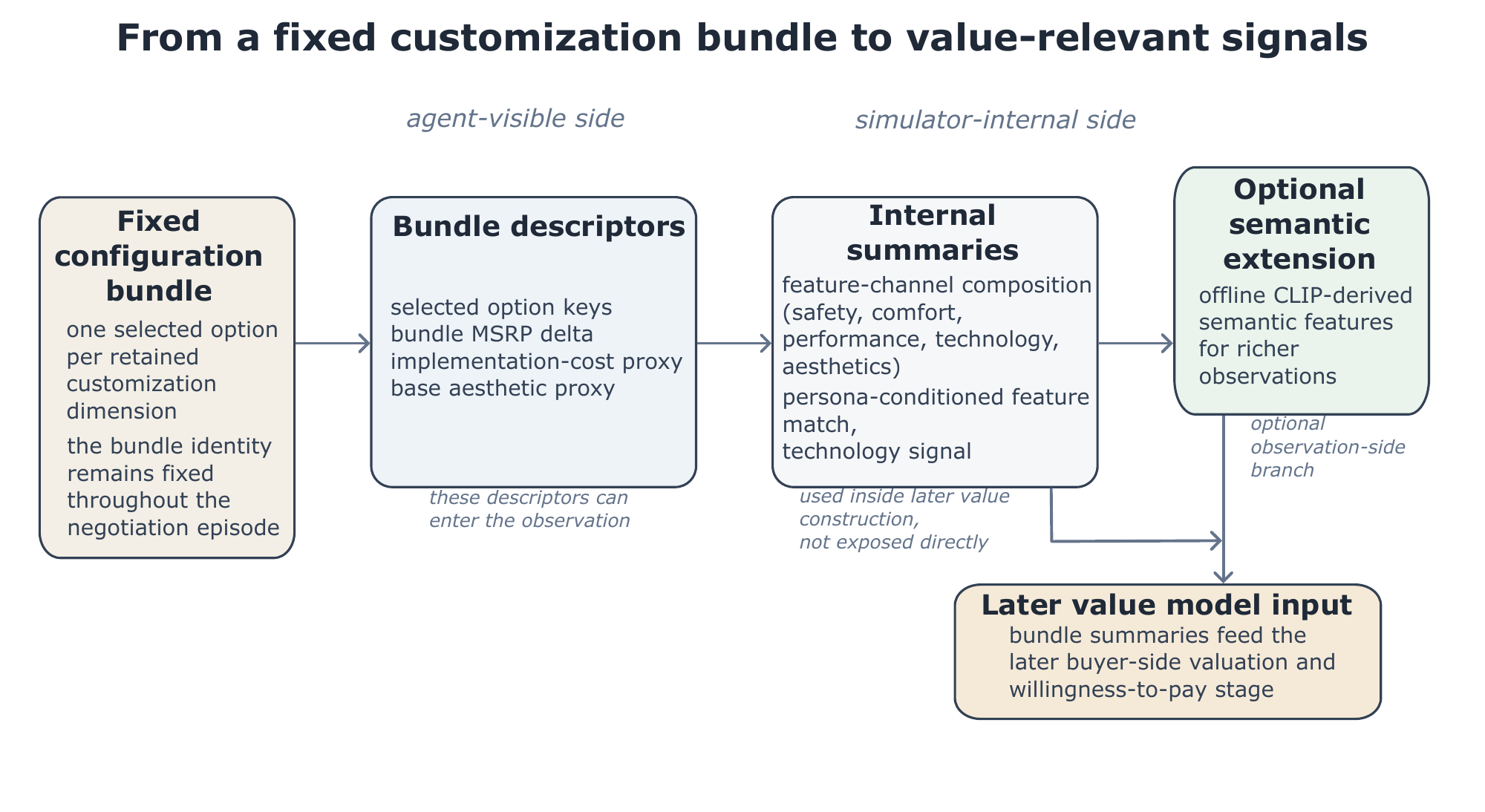}
  \caption{Bundle-signal construction. Fixed customization descriptors are visible to the seller, while value-relevant summaries are computed inside the simulator and remain hidden from the agent.}
  \label{fig:bundle_signals}
\end{figure}

The simulator converts each selected bundle into valuation signals through a staged representation. \Cref{fig:bundle_signals} illustrates this conversion. The first layer contains seller-visible bundle descriptors: selected option keys, dimensions, total MSRP delta, estimated implementation cost, and aesthetic proxy. The second layer computes simulator-internal bundle summaries, including a feature-channel composition over safety, comfort, performance, technology, and aesthetics. The final layer combines those internal bundle summaries with hidden persona preferences to form persona-conditioned signals such as feature match and technology value. This representation gives the seller an interpretable bundle summary while keeping preference-matched valuation features inside the simulator.

Given a persona and bundle, the simulator computes round-level willingness to pay as
\begin{equation}
  \mathrm{WTP}_t = \max\left\{1000,\,
  R_{\mathrm{base}} + V_{\mathrm{custom}} + V_{\mathrm{aesthetic}} + V_{\mathrm{brand\text{-}tech}} - V_{\mathrm{fatigue}} + \epsilon_t
  \right\}.
\label{eq:wtp}
\end{equation}
The baseline term \(R_{\mathrm{base}}\) reflects hidden reservation-price level and price sensitivity. The customization term \(V_{\mathrm{custom}}\) uses the bundle's MSRP anchor and persona-conditioned feature match. The aesthetic and brand--technology terms capture hidden taste for appearance, brand, and technology-heavy bundles. The fatigue term reduces effective valuation as bargaining progresses, with the strength of this effect shaped by hidden patience and impulsivity. The noise term \(\epsilon_t\) adds round-level stochasticity, while the lower bound keeps willingness to pay at a small positive floor.

Buyer response is generated from buyer-side utility. For a seller offer \(p_t\), buyer utility is
\begin{equation}
  U^{\mathrm{buyer}}_t(p_t) = \mathrm{WTP}_t - p_t .
\label{eq:buyer_utility}
\end{equation}
When \(U^{\mathrm{buyer}}_t(p_t) \geq 0\), the simulated buyer accepts the offer. When \(U^{\mathrm{buyer}}_t(p_t) < 0\), the simulator computes a utility-gap-dependent walkaway probability shaped by hidden price sensitivity and walkaway threshold; if the buyer does not walk away, the session continues through rejection or a counter-offer whose level is influenced by counter strength. This mechanism makes buyer behavior depend jointly on the fixed bundle, latent persona structure, round context, and current price.

PrefBench realizes the interaction through a persistent NegMAS bargaining session \citep{mohammad2021NegMASPlatformAutomated}. Each episode creates one session for the sampled persona and fixed bundle. Seller offers update the continuing session, and the simulated buyer accepts, rejects, counters, or exits according to the current buyer utility and hidden bargaining variables. If the episode continues, the latest offer, response, counter-offer, and session history carry forward to the next seller decision. This persistent session preserves continuity across rounds while keeping price as the only negotiated issue.

\section{LLM Protocol and Baselines}
\label{sec:llm_protocol}

\begin{figure}[t]
  \centering
  \includegraphics[width=0.98\linewidth]{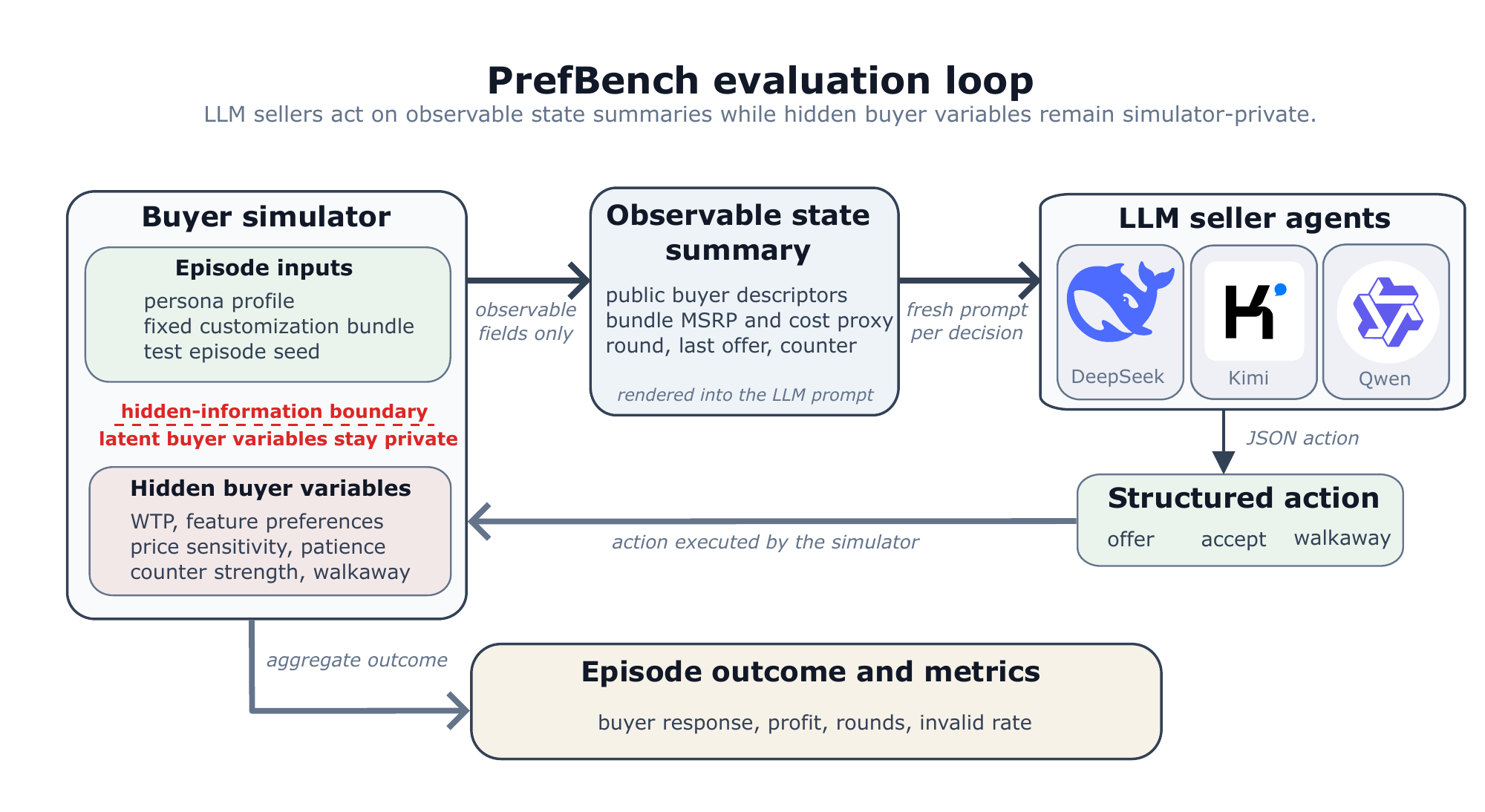}
  \caption{LLM-facing PrefBench evaluation loop. The LLM seller receives an observable state summary, returns one structured JSON action, and is evaluated through simulator outcomes while hidden buyer variables remain private to the simulator.}
  \label{fig:llm_protocol_architecture}
\end{figure}

LLM agents are evaluated as prompt-based seller policies under the same hidden-information boundary as the heuristic references. \Cref{fig:llm_protocol_architecture} shows the interaction contract. At each seller decision point, the agent receives a text rendering of the observable state and returns one structured JSON action. The protocol follows the standard OpenAI-compatible chat-completions format \citep{openai2026ChatCompletionsAPI} and maps each returned object to the shared \texttt{EnvAction} schema used by the simulator. This interface places LLM and non-LLM policies in the same environment while exposing LLMs only to a text-rendered state summary.

The LLM policy uses decision-level prompting with history encoded in the observable state summary. Each seller decision is produced from the current environment observation through a separate model call. The prompt includes the current round, remaining seller turns, the last seller offer, the latest buyer response, the latest buyer counter-offer if one exists, and the negotiation history length. This design makes multi-round context explicit and reproducible while avoiding dependence on an unbounded chat transcript. The complete prompt template is provided in \cref{app:llm_prompt_template}.

The rendered observation follows the information boundary defined in \cref{sec:task_formulation}. It exposes the selected customization options, bundle MSRP delta, estimated implementation cost, aesthetic proxy, observable buyer profile fields, and the interaction-state summary. Simulator-side variables remain hidden, including true willingness to pay, reservation-price level, price sensitivity, feature-preference weights, patience, counter strength, and walkaway threshold. The prompt also states the seller objective and benchmark profit formula, emphasizing that the objective distinguishes agreement from seller profit.

The output contract defines both the action schema and the evaluation accounting. The prompt requests one JSON object with a \texttt{move}, a \texttt{price\_offer\_usd}, and a brief \texttt{reason}. A valid action requires one of three supported moves: \texttt{offer}, \texttt{accept}, or \texttt{walkaway}. An \texttt{offer} additionally requires a non-negative numeric price; \texttt{accept} applies only when the buyer has made a counter-offer; and \texttt{reason} is recorded for trace inspection rather than scoring. Parser-invalid outputs include malformed JSON, non-object responses, unsupported moves, missing offer prices, non-numeric offer prices, and negative offer prices. Schema-valid actions whose environmental preconditions are unsatisfied, such as \texttt{accept} before a buyer counter-offer exists, are recorded as unavailable environment actions. Invalid outputs terminate the current episode without a replacement action, so the invalid-output rate directly measures structured-output compliance.

The main prompt version, denoted prompt v1, gives the model the role of seller, states the profit objective, describes the hidden-information limit, defines the action meanings, and provides short decision guidance such as avoiding below-cost offers and balancing agreement with seller profit. A longer prompt v2 adds more field-level explanation and serves as a prompt-clarity ablation in \cref{sec:experiments}. The main experiments use prompt v1 for all reported LLM baselines.

The comparison includes two non-learned heuristic references and three zero-shot LLM baselines. Here, zero-shot denotes instruction-only prompting: the models receive the task description, state summary, objective, and output contract, but no task-specific training, fine-tuning, reinforcement learning, or in-context demonstration episodes. The random and concession heuristics consume structured observations and provide non-learned reference policies. The LLM baselines are DeepSeek V4 Flash \citep{deepseekai2026deepseekv4}, Kimi K2.6 \citep{moonshot2026KimiK26Blog}, and Qwen3.6 Plus \citep{qwen36plus}, all called through provider-specific OpenAI-compatible endpoints. We include a DeepSeek V4 Flash prompt-v2 ablation and a 100-episode DeepSeek V4 Pro reasoning-enabled run as secondary analyses. \Cref{tab:llm_setup} summarizes the model and decoding settings.

\begin{table}[t]
  \centering
  \caption{LLM run configuration. All rows use OpenAI-compatible chat-completions calls; model documentation and provider pages were accessed in May 2026.}
  \label{tab:llm_setup}
  \scriptsize
  \setlength{\tabcolsep}{2.5pt}
  \begin{tabular*}{\linewidth}{@{\extracolsep{\fill}}lllccclr@{}}
    \toprule
    Provider & Run & API model ID & Prompt & Temp. & Max & Output / reasoning & Episodes \\
    \midrule
    DeepSeek & V4 Flash & \texttt{deepseek-v4-flash} & v1 & 0.0 & 512 & JSON; thinking disabled & 7,500 \\
    Moonshot & Kimi K2.6 & \texttt{kimi-k2.6} & v1 & 0.6 & 512 & JSON; thinking disabled & 7,500 \\
    Alibaba & Qwen3.6 Plus & \texttt{qwen3.6-plus} & v1 & 0.0 & 512 & JSON; thinking disabled & 7,500 \\
    DeepSeek & V4 Flash ablation & \texttt{deepseek-v4-flash} & v2 & 0.0 & 512 & JSON; thinking disabled & 7,500 \\
    DeepSeek & V4 Pro reasoning & \texttt{deepseek-v4-pro} & v1 & 0.0 & 2048 & JSON; thinking enabled & 100 \\
    \bottomrule
  \end{tabular*}
\end{table}

\section{Experiments}
\label{sec:experiments}

The main experiments use a fixed 7,500-episode held-out evaluation split generated with seed 123. Across all methods, the environment, hidden-information boundary, action space, and episode stream are held constant. Heuristic policies act on structured observations, while LLM policies act on prompt-rendered state summaries.

The two heuristic policies provide transparent reference points for the fixed protocol. The random policy samples legal offers and occasionally accepts available buyer counters or walks away. The concession heuristic starts from an anchor of approximately \(2.2\) times the bundle MSRP delta, concedes linearly toward a floor of \(1.1\) times the bundle MSRP delta, and conditions counter-offer acceptance on that floor. \Cref{app:experiment_uncertainty} gives the full heuristic specification.

\Cref{tab:main_results} summarizes the main comparison. Parenthesized values are changes relative to the random policy. The primary metric is average realized seller profit; deal rate, profit per deal, average rounds, and invalid rate diagnose how that profit is obtained. Profit per deal is computed from exact deal counts and total realized profit before rounding. Invalid rate measures parser-level output validity for LLM runs. \Cref{app:experiment_uncertainty} provides bootstrap confidence intervals for deal rate and average profit.

\begin{table}[t]
  \centering
  \caption{Main PrefBench results on the 7,500-episode test split.}
  \label{tab:main_results}
  \small
  \setlength{\tabcolsep}{2pt}
  \begin{tabular}{@{}lrrrrr@{}}
    \toprule
    Method & Deal Rate \(\uparrow\) & Avg Profit \(\uparrow\) & Profit / Deal \(\uparrow\) & Avg Rounds & Invalid \(\downarrow\) \\
    \midrule
    Random & 0.5769 & 6,572.33 & 11,391.84 & 1.3541 & n/a \\
    Concession & 0.7268 {\color{green!50!black}(+0.1499)} & \textbf{14,774.11} {\color{green!50!black}(+8,201.78)} & \textbf{20,327.62} & 1.7123 & n/a \\
    DeepSeek V4 Flash~\citep{deepseekai2026deepseekv4} & 0.9903 {\color{green!50!black}(+0.4133)} & 6,749.21 {\color{green!50!black}(+176.87)} & 6,815.55 & 1.0313 & 0.0000 \\
    Kimi K2.6~\citep{moonshot2026KimiK26Blog} & \textbf{1.0000} {\color{green!50!black}(+0.4231)} & 4,514.43 {\color{red!70!black}(-2,057.90)} & 4,514.43 & \textbf{1.0000} & 0.0000 \\
    Qwen3.6 Plus~\citep{qwen36plus} & 0.9985 {\color{green!50!black}(+0.4216)} & 5,979.55 {\color{red!70!black}(-592.79)} & 5,988.33 & 1.0036 & 0.0000 \\
    \bottomrule
  \end{tabular}
\end{table}

The main table separates structured validity, deal rate, and deal quality. All three LLM baselines produce valid actions throughout the evaluation split and reach deal rates above 0.99. The concession heuristic achieves the highest seller returns, with 14,774.11 USD average profit and 20,327.62 USD profit per deal. The strongest LLM result on average profit is DeepSeek V4 Flash at 6,749.21 USD, while Kimi K2.6 and Qwen3.6 Plus fall below the random baseline despite closing more deals.

The round statistics indicate that LLM agreements are typically immediate. A one-round episode denotes termination after the first seller decision. Under this definition, DeepSeek V4 Flash, Kimi K2.6, and Qwen3.6 Plus terminate in the first round in 7,408, 7,500, and 7,489 of the 7,500 episodes, respectively. \Cref{sec:discussion} interprets these result patterns.

\Cref{tab:supporting_ablations} reports two supporting LLM analyses. The prompt-v2 run keeps the same action schema as prompt v1 but adds explicit scenario, evaluation, interaction-dynamics, and state-description fields; \cref{app:llm_prompt_template} records the exact prompt difference. Compared with DeepSeek V4 Flash prompt v1, prompt v2 increases deal rate from 0.9903 to 0.9993 (\(+0.0090\)) but decreases average profit from 6,749.21 USD to 4,142.73 USD (\(-2{,}606.48\) USD). The 100-episode DeepSeek V4 Pro reasoning run reaches a 0.9900 deal rate and 5,231.50 USD average profit with one parser-invalid output.

\begin{table}[t]
  \centering
  \caption{Supporting LLM analyses. Prompt v2 is run on the 7,500-episode split; the reasoning run uses 100 episodes.}
  \label{tab:supporting_ablations}
  \small
  \setlength{\tabcolsep}{4pt}
  \begin{tabular}{@{}lrrrr@{}}
    \toprule
    Run & Deal Rate \(\uparrow\) & Avg Profit \(\uparrow\) & Avg Rounds & Invalid \(\downarrow\) \\
    \midrule
    DeepSeek V4 Flash prompt v2~\citep{deepseekai2026deepseekv4} & 0.9993 & 4,142.73 & 1.0008 & 0.0000 \\
    DeepSeek V4 Pro Reasoning~\citep{deepseekai2026deepseekv4} & 0.9900 & 5,231.50 & 1.0000 & 0.0100 \\
    \bottomrule
  \end{tabular}
\end{table}

\section{Findings and Discussion}
\label{sec:discussion}

The main finding is that structured action compliance and strategic pricing competence are separable. In the main full-test runs, the tested LLMs follow the JSON action interface reliably and produce no parser-invalid episodes. This establishes that the task can be exposed through a simple structured action contract. At the same time, action validity alone does not indicate profit-sensitive bargaining competence.

The LLM policies are strongly biased toward immediate settlement. DeepSeek V4 Flash, Kimi K2.6, and Qwen3.6 Plus terminate in the first round in 7,408, 7,500, and 7,489 of the 7,500 test episodes. The prompt exposes the current round, remaining seller turns, the latest offer state, and an instruction to avoid optimizing only for immediate acceptance, yet the resulting policies still concentrate on first-round agreement. This pattern suggests that, under zero-shot prompting, the models prioritize acceptance probability over seller surplus when buyer willingness-to-pay is hidden. In effect, they convert the task into a one-shot agreement problem rather than a multi-round price-discovery problem.

High deal rate consequently does not imply high seller profit. The LLM baselines reach deal rates above 0.99, but their average profits remain far below the concession heuristic. The concession policy closes fewer deals and accepts more walkaway risk, yet it earns substantially higher average profit because it encodes an explicit aggressive-concession prior. The heuristic therefore serves as a diagnostic reference: it shows that the simulator rewards strategic anchoring and concession behavior, and that the tested zero-shot LLMs do not discover that behavior from the prompt alone.

The result also clarifies what the benchmark measures. A model can produce valid structured actions and maximize apparent buyer agreement while still performing poorly on the seller objective. This distinction matters for LLM-agent evaluation because format compliance is often a prerequisite for structured agent interaction, but it is not a sufficient proxy for decision quality. PrefBench makes this gap observable by separating three quantities that are easily conflated: structured action compliance, agreement-seeking behavior, and profit-sensitive decision quality.

The prompt and reasoning ablations reinforce the same interpretation. A more detailed prompt did not improve DeepSeek V4 Flash's strategy and instead made the model more conservative, increasing deal rate while lowering average profit. In this ablation, prompt v2 mainly tests whether more explicit scenario and state-field descriptions change the bargaining pattern while the action schema remains fixed; the full prompt is provided in \cref{app:llm_prompt_template}. The 100-episode reasoning-enabled run also does not produce multi-round bargaining behavior. Because these ablations are limited in scope, they are best interpreted as diagnostic evidence: hidden-preference pricing appears to require more than longer task descriptions or generic reasoning mode. The most direct next directions are explicit value estimation, planning over latent buyer types, and policy learning from simulated trajectories.

Overall, the results position PrefBench as a diagnostic benchmark for studying pricing-agent behavior under hidden buyer preferences. The tested models can use the action protocol reliably, but the benchmark exposes a decision-making failure mode: agreement-oriented behavior when buyer valuation remains hidden. This failure mode is difficult to isolate in human-facing or open-domain pricing studies, but it becomes measurable in a controlled simulator with fixed hidden variables, repeated episode streams, and explicit seller-profit metrics.

\section{Limitations and Future Work}
\label{sec:limitations}

PrefBench uses a controlled simulator to isolate hidden-preference pricing behavior under a fixed protocol. This design supports reproducible comparison while leaving external validity as a separate question. The buyer population is public-data-informed but semi-synthetic, and the hidden preference and response model is benchmark-defined; real negotiation traces would be needed to estimate these mechanisms empirically. The profit values measure simulator-level seller outcomes, not real-world revenue. Future work should validate selected parts of the simulator against human buyer or seller responses, including acceptance, counter-offer, and walkaway behavior.

The evaluation covers one product substrate, one main prompt protocol, and a selected set of hosted LLMs. LLM behavior may vary with different product domains, prompts, provider settings, and model-version changes. The reasoning-enabled evaluation is limited to 100 episodes. A broader evaluation should include fine-tuned LLMs, learned pricing policies, tool-augmented agents, session-based LLM agents, and reasoning-model families under matched cost budgets. These extensions are important for separating zero-shot prompt behavior under the PrefBench protocol from broader pricing-agent capability.

PrefBench focuses on seller-side decision behavior; welfare, fairness, privacy, and policy analysis remain outside its current scope. A broader evaluation would add metrics for consumer surplus, fairness across observable groups, robustness to profile shifts, inventory effects, and long-term customer value. Such extensions would test whether pricing agents can maintain profitability while treating seller profit as one objective among several consumer-facing constraints.

PrefBench is one focused benchmark instance within a broader pricing-agent evaluation suite. The task studies hidden-preference negotiation over a fixed sampled customization bundle. Natural extensions include agents that choose or revise bundles, incorporate inventory and implementation-cost constraints, and reason across repeated customer interactions. More broadly, pricing-agent benchmarks should cover multiple pricing mechanisms and consumer-facing objectives; PrefBench provides a reproducible starting point for that broader evaluation direction.

\section{Conclusion}
\label{sec:conclusion}

This paper presents PrefBench, a controlled simulator-based benchmark for evaluating seller-side LLM agents in hidden-preference personalized pricing negotiations. The benchmark places agents in a fixed hidden-information setting with latent buyer variables, a seller-side profit objective, and a structured JSON action protocol. This setting tests whether valid structured interaction translates into profit-sensitive pricing behavior.

The zero-shot LLM results show that the tested models can follow the protocol and reach agreements reliably, but high action compliance and high deal rate do not imply profit-sensitive bargaining. The models tend to settle immediately and earn substantially lower seller profit than a simple concession heuristic under the same episode stream. PrefBench offers a focused evaluation setting for separating action-format compliance, agreement-seeking behavior, and strategic pricing quality in pricing agents under hidden buyer preferences.

\bibliographystyle{unsrtnat}
\bibliography{data/UG_Thesis}

\begin{thebibliography}{26}
\providecommand{\natexlab}[1]{#1}
\providecommand{\url}[1]{\texttt{#1}}
\expandafter\ifx\csname urlstyle\endcsname\relax
  \providecommand{\doi}[1]{doi: #1}\else
  \providecommand{\doi}{doi: \begingroup \urlstyle{rm}\Url}\fi

\bibitem[Choi et~al.(2023)Choi, Kim, Choi, Cho, Paik, and
  Oh]{choi2023SemiParametricContextualPricing}
Young-Geun Choi, Gi-Soo Kim, Yunseo Choi, Wooseong Cho, Myunghee~Cho Paik, and
  Min-Hwan Oh.
\newblock Semi-{{Parametric Contextual Pricing Algorithm}} using {{Cox
  Proportional Hazards Model}}.
\newblock In \emph{Proceedings of the 40th {{International Conference}} on
  {{Machine Learning}}}, pages 5771--5786. PMLR, July 2023.

\bibitem[Liu et~al.(2021)Liu, Zhang, Wang, Deng, and
  Wu]{liu2021DynamicPricingEcommerce}
Jiaxi Liu, Yidong Zhang, Xiaoqing Wang, Yuming Deng, and Xingyu Wu.
\newblock Dynamic {{Pricing}} on {{E-commerce Platform}} with {{Deep
  Reinforcement Learning}}: {{A Field Experiment}}.
\newblock Technical Report arXiv:1912.02572, arXiv, August 2021.

\bibitem[Biggs et~al.(2021)Biggs, Sun, and
  Ettl]{biggs2021ModelDistillationRevenue}
Max Biggs, Wei Sun, and Markus Ettl.
\newblock Model distillation for revenue optimization: Interpretable
  personalized pricing.
\newblock In \emph{International Conference on Machine Learning}, pages
  946--956. PMLR, 2021.

\bibitem[Dub{\'e} and Misra(2023)]{dube2017PersonalizedPricingConsumer}
Jean-Pierre Dub{\'e} and Sanjog Misra.
\newblock Personalized pricing and consumer welfare.
\newblock \emph{Journal of Political Economy}, 131\penalty0 (1):\penalty0
  131--189, 2023.

\bibitem[Xia et~al.(2023)Xia, Arian, Narayanamoorthy, and
  Mabry]{xia2023RetailSynthSyntheticData}
Yu~Xia, Ali Arian, Sriram Narayanamoorthy, and Joshua Mabry.
\newblock {{RetailSynth}}: {{Synthetic Data Generation}} for {{Retail AI
  Systems Evaluation}}.
\newblock Technical Report arXiv:2312.14095, arXiv, December 2023.

\bibitem[Baarslag et~al.(2012)Baarslag, Hindriks, Jonker, Kraus, and
  Lin]{baarslag2012FirstAutomatedNegotiating}
Tim Baarslag, Koen Hindriks, Catholijn Jonker, Sarit Kraus, and Raz Lin.
\newblock The {{First Automated Negotiating Agents Competition}} ({{ANAC}}
  2010).
\newblock In Takayuki Ito, Minjie Zhang, Valentin Robu, Shaheen Fatima, and
  Tokuro Matsuo, editors, \emph{New {{Trends}} in {{Agent-Based Complex
  Automated Negotiations}}}, pages 113--135. Springer, Berlin, Heidelberg,
  2012.
\newblock ISBN 978-3-642-24696-8.
\newblock \doi{10.1007/978-3-642-24696-8_7}.

\bibitem[Lin et~al.(2014)Lin, Kraus, Baarslag, Tykhonov, Hindriks, and
  Jonker]{lin2014GeniusIntegratedEnvironment}
Raz Lin, Sarit Kraus, Tim Baarslag, Dmytro Tykhonov, Koen Hindriks, and
  Catholijn~M. Jonker.
\newblock Genius: {{An Integrated Environment}} for {{Supporting}} the
  {{Design}} of {{Generic Automated Negotiators}}.
\newblock \emph{Computational Intelligence}, 30\penalty0 (1):\penalty0 48--70,
  2014.
\newblock ISSN 1467-8640.
\newblock \doi{10.1111/j.1467-8640.2012.00463.x}.

\bibitem[Xia et~al.(2024)Xia, He, Ren, Miao, Zhang, Yang, and
  Wang]{xia2024MeasuringBargainingAbilities}
Tian Xia, Zhiwei He, Tong Ren, Yibo Miao, Zhuosheng Zhang, Yang Yang, and Rui
  Wang.
\newblock Measuring bargaining abilities of llms: A benchmark and a
  buyer-enhancement method.
\newblock In \emph{Findings of the Association for Computational Linguistics:
  ACL 2024}, pages 3579--3602, 2024.

\bibitem[Chan et~al.(2024)Chan, Cheng, Yim, Deng, Fan, Li, Liu, Zhang, Wang,
  and Song]{chan2024NegotiationToMBenchmarkStresstesting}
Chunkit Chan, Jiayang Cheng, Yauwai Yim, Zheye Deng, Wei Fan, Haoran Li, Xin
  Liu, Hongming Zhang, Weiqi Wang, and Yangqiu Song.
\newblock Negotiationtom: A benchmark for stress-testing machine theory of mind
  on negotiation surrounding.
\newblock In \emph{Findings of the Association for Computational Linguistics:
  EMNLP 2024}, pages 4211--4241, 2024.

\bibitem[Liu et~al.(2023)Liu, Yu, Zhang, Xu, Lei, Lai, Gu, Ding, Men, Yang,
  Zhang, Deng, Zeng, Du, Zhang, Shen, Zhang, Su, Sun, Huang, Dong, and
  Tang]{liu2023AgentBenchEvaluatingLLMs}
Xiao Liu, Hao Yu, Hanchen Zhang, Yifan Xu, Xuanyu Lei, Hanyu Lai, Yu~Gu,
  Hangliang Ding, Kaiwen Men, Kejuan Yang, Shudan Zhang, Xiang Deng, Aohan
  Zeng, Zhengxiao Du, Chenhui Zhang, Sheng Shen, Tianjun Zhang, Yu~Su, Huan
  Sun, Minlie Huang, Yuxiao Dong, and Jie Tang.
\newblock {{AgentBench}}: {{Evaluating LLMs}} as {{Agents}}, 2023.
\newblock URL \url{https://arxiv.org/abs/2308.03688}.

\bibitem[Qin et~al.(2023)Qin, Liang, Ye, Zhu, Yan, Lu, Lin, Cong, Tang, Qian,
  Zhao, Hong, Tian, Xie, Zhou, Gerstein, Li, Liu, and
  Sun]{qin2023ToolLLMFacilitatingLarge}
Yujia Qin, Shihao Liang, Yining Ye, Kunlun Zhu, Lan Yan, Yaxi Lu, Yankai Lin,
  Xin Cong, Xiangru Tang, Bill Qian, Sihan Zhao, Lauren Hong, Runchu Tian,
  Ruobing Xie, Jie Zhou, Mark Gerstein, Dahai Li, Zhiyuan Liu, and Maosong Sun.
\newblock {{ToolLLM}}: {{Facilitating Large Language Models}} to {{Master}}
  16000+ {{Real-World APIs}}, 2023.
\newblock URL \url{https://arxiv.org/abs/2307.16789}.

\bibitem[Yao et~al.(2024)Yao, Shinn, Razavi, and
  Narasimhan]{yao2024TauBenchToolAgentUser}
Shunyu Yao, Noah Shinn, Pedram Razavi, and Karthik Narasimhan.
\newblock {$\tau$-Bench}: {{A Benchmark}} for {{Tool-Agent-User Interaction}}
  in {{Real-World Domains}}, 2024.
\newblock URL \url{https://arxiv.org/abs/2406.12045}.

\bibitem[Pandey et~al.(2020)Pandey, Wang, and
  Boyles]{pandey2020DeepReinforcementLearning}
Venktesh Pandey, Evana Wang, and Stephen~D. Boyles.
\newblock Deep {{Reinforcement Learning Algorithm}} for {{Dynamic Pricing}} of
  {{Express Lanes}} with {{Multiple Access Locations}}.
\newblock \emph{Transportation Research Part C: Emerging Technologies},
  119:\penalty0 102715, October 2020.
\newblock ISSN 0968090X.
\newblock \doi{10.1016/j.trc.2020.102715}.

\bibitem[Priester et~al.(2020)Priester, Robbert, and
  Roth]{priester2020SpecialPriceJust}
Anna Priester, Thomas Robbert, and Stefan Roth.
\newblock A special price just for you: Effects of personalized dynamic pricing
  on consumer fairness perceptions.
\newblock \emph{Journal of Revenue and Pricing Management}, 19\penalty0
  (2):\penalty0 99--112, April 2020.
\newblock ISSN 1477-657X.
\newblock \doi{10.1057/s41272-019-00224-3}.

\bibitem[Thomassen(2017)]{thomassen2017EmpiricalModelAutomobile}
{\O}yvind Thomassen.
\newblock An {{Empirical Model}} of {{Automobile Engine Variant Pricing}}.
\newblock \emph{International Journal of the Economics of Business},
  24\penalty0 (3):\penalty0 275--293, September 2017.
\newblock ISSN 1357-1516.
\newblock \doi{10.1080/13571516.2017.1333733}.

\bibitem[Wang et~al.(2022)Wang, Chen, and
  Wang]{wang2022AssortmentPlanningPricing}
Yana Wang, Zhen-Song Chen, and Xian-Jia Wang.
\newblock Assortment planning and pricing for configurable product under
  sequential choice process.
\newblock \emph{Management System Engineering}, 1\penalty0 (1):\penalty0 6,
  October 2022.
\newblock ISSN 2731-5843.
\newblock \doi{10.1007/s44176-022-00002-3}.

\bibitem[Mohammad et~al.(2021)Mohammad, Nakadai, and
  Greenwald]{mohammad2021NegMASPlatformAutomated}
Yasser Mohammad, Shinji Nakadai, and Amy Greenwald.
\newblock {{NegMAS}}: {{A Platform}} for {{Automated Negotiations}}.
\newblock In Takahiro Uchiya, Quan Bai, and Iv{\'a}n Mars{\'a}~Maestre,
  editors, \emph{{{PRIMA}} 2020: {{Principles}} and {{Practice}} of
  {{Multi-Agent Systems}}}, volume 12568, pages 343--351. Springer
  International Publishing, Cham, 2021.
\newblock ISBN 978-3-030-69321-3 978-3-030-69322-0.
\newblock \doi{10.1007/978-3-030-69322-0_23}.

\bibitem[{Villarrubia-Martin} et~al.(2025){Villarrubia-Martin},
  {Rodriguez-Benitez}, {Mu{\~n}oz-Valero}, Montana, and
  {Jimenez-Linares}]{villarrubia-martin2025DynamicPricingHighSpeed}
Enrique~Adrian {Villarrubia-Martin}, Luis {Rodriguez-Benitez}, David
  {Mu{\~n}oz-Valero}, Giovanni Montana, and Luis {Jimenez-Linares}.
\newblock Dynamic {{Pricing}} in {{High-Speed Railways Using Multi-Agent
  Reinforcement Learning}}.
\newblock Technical Report arXiv:2501.08234, arXiv, September 2025.

\bibitem[{Census Reporter}(2026)]{CensusProfileUnited}
{Census Reporter}.
\newblock Census profile: {{United States}}.
\newblock http://censusreporter.org/profiles/01000US-united-states/, 2026.

\bibitem[{U.S. Census Bureau}(2024)]{bureauIncomeUnitedStates}
{U.S. Census Bureau}.
\newblock Income in the {{United States}}: 2023.
\newblock https://www.census.gov/library/publications/2024/demo/p60-282.html,
  2024.

\bibitem[Bricka et~al.(2024)Bricka, Reuscher, Schroeder, Fisher, Beard, and
  Sun]{bricka2024summary}
Stacey Bricka, Timothy Reuscher, Paul Schroeder, Mitchell Fisher, Justina
  Beard, and Xiaoyuan~Layla Sun.
\newblock Summary of travel trends: 2022 national household travel survey.
\newblock Technical report, Federal Highway Administration, 2024.

\bibitem[{Mercedes-Benz USA}(2026)]{BuildYourOwn}
{Mercedes-Benz USA}.
\newblock Build {{Your Own}} 2026 {{E}} 350 {{Sedan}}.
\newblock https://www.mbusa.com/en/vehicles/build/e-class/sedan/e350w, 2026.

\bibitem[{OpenAI}(2026)]{openai2026ChatCompletionsAPI}
{OpenAI}.
\newblock {{Chat Completions}}.
\newblock OpenAI API Reference, 2026.
\newblock URL \url{https://developers.openai.com/api/reference/resources/chat}.

\bibitem[{DeepSeek-AI}(2026)]{deepseekai2026deepseekv4}
{DeepSeek-AI}.
\newblock {{DeepSeek-V4}}: {{Towards Highly Efficient Million-Token Context
  Intelligence}}, 2026.
\newblock URL \url{https://huggingface.co/collections/deepseek-ai/deepseek-v4}.

\bibitem[{Moonshot AI}(2026)]{moonshot2026KimiK26Blog}
{Moonshot AI}.
\newblock {{Kimi K2.6}}: {{Advancing Open-Source Coding}}.
\newblock Kimi Technical Blog, 2026.
\newblock URL \url{https://www.kimi.com/blog/kimi-k2-6}.

\bibitem[{Qwen Team}(2026)]{qwen36plus}
{Qwen Team}.
\newblock {{Qwen3.6-Plus}}: {{Towards Real World Agents}}, April 2026.
\newblock URL \url{https://qwen.ai/blog?id=qwen3.6}.

\end{thebibliography}

\clearpage
\appendix
\section{Customization Scope}
\label{app:customization_scope}

PrefBench uses a focused Mercedes-Benz E350 Sedan customization catalog as the fixed product substrate. The catalog is derived from selected official configuration options and MSRP deltas \citep{BuildYourOwn}, then standardized into canonical benchmark options. The scope is defined to capture representative customization choices relevant to pricing and negotiation. It retains high-impact options and abstracts away fine-grained accessories and package-dependency rules.

Each episode draws one option from each retained customization dimension. Single-option dimensions remain fixed across sampled bundles. The selected options form a fixed bundle that remains unchanged throughout the negotiation. The decision problem is price negotiation over this fixed bundle; bundle design and product recommendation are outside the appendix-defined task scope. MSRP deltas are consumer-facing option deltas, while the seller-side implementation cost used in the benchmark is a fixed proxy derived from the selected bundle.

\begin{table}[H]
  \centering
  \caption{E350 customization catalog used to define the PrefBench product scope. MSRP deltas are consumer-facing option deltas compiled into the benchmark catalog.}
  \label{tab:appendix_customization_catalog}
  \scriptsize
  \setlength{\tabcolsep}{4pt}
  \begin{tabular}{@{}p{0.12\linewidth}p{0.24\linewidth}p{0.40\linewidth}rr@{}}
    \toprule
    Dimension & Option ID & Representative option(s) & MSRP & Aesthetic \\
    \midrule
    Paint color & \texttt{paint\_standard} & Standard black or Polar White paint & 0 & 0.20 \\
    Paint color & \texttt{paint\_metallic} & Metallic paint finish & 750 & 0.45 \\
    Paint color & \texttt{paint\_manufaktur} & MANUFAKTUR paint finish & 1,750 & 0.80 \\
    Wheels & \texttt{wheel\_18\_standard} & 18-inch standard wheel set & 0 & 0.20 \\
    Wheels & \texttt{wheel\_19\_upgrade} & 19-inch wheel upgrade & 600 & 0.50 \\
    Wheels & \texttt{wheel\_amg\_high} & AMG 20/21-inch wheel upgrade & 1,950 & 0.85 \\
    Exterior style & \texttt{styling\_upgrade} & Night Package or illuminated grille styling upgrade & 400 & 0.55 \\
    Upholstery & \texttt{mb\_tex} & MB-Tex upholstery & 0 & 0.25 \\
    Upholstery & \texttt{leather} & Leather upholstery & 1,620 & 0.65 \\
    Upholstery & \texttt{nappa\_leather} & Nappa leather upholstery & 2,990 & 0.90 \\
    Trim & \texttt{standard\_trim} & Standard wood or piano-black trim & 0 & 0.25 \\
    Trim & \texttt{premium\_trim} & Premium wood, metallic, or star-pattern trim & 150 & 0.55 \\
    Comfort & \texttt{multicontour\_package} & Multicontour Seating Package & 2,950 & 0.85 \\
    Comfort & \texttt{seat\_comfort\_upgrade} & Ventilated front seats or heated rear seats & 500 & 0.45 \\
    Comfort & \texttt{soft\_close\_doors} & Soft-close doors & 550 & 0.40 \\
    Audio & \texttt{burmester\_4d} & Burmester 4D surround-sound system & 1,030 & 0.70 \\
    Technology & \texttt{mbux\_superscreen} & MBUX Superscreen Package & 1,500 & 0.90 \\
    Safety & \begin{tabular}[t]{@{}l@{}}\texttt{driver\_assistance}\\\texttt{\_package}\end{tabular} & Driver Assistance Package & 1,950 & 0.60 \\
    Performance & \texttt{airmatic\_package} & AIRMATIC Package & 3,200 & 0.65 \\
    Lighting & \texttt{digital\_light} & DIGITAL LIGHT headlamps & 990 & 0.60 \\
    \bottomrule
  \end{tabular}
\end{table}

\clearpage
\section{Persona Priors and Conditional Distributions}
\label{app:persona_conditionals}

This appendix documents the probability tables and generation rules used by the persona schema. The observable profile is informed by public U.S. demographic, income, and travel-purpose sources together with lightweight conditional approximations, while the hidden layer is benchmark-defined through conditional mappings, numeric mixtures, bounded shifts, and coupling rules. These tables expose benchmark assumptions for auditability; they are design parameters for controlled persona heterogeneity, not empirical estimates of individual buyer parameters.

\subsection{Observable Priors}

\begin{table}[H]
\centering
\small
\caption{Observable profile priors used by the consumer simulator.}
\label{tab:observable_priors}
\setlength{\tabcolsep}{4pt}
\begin{tabular}{@{}p{0.18\linewidth}p{0.34\linewidth}p{0.38\linewidth}@{}}
\toprule
Field & Values & Probability vector \\
\midrule
Age band & 18--25, 26--35, 36--50, 50+ & [0.08, 0.24, 0.39, 0.29] \\
Income band & <60k, 60--100k, 100--180k, 180k+ & [0.08, 0.22, 0.40, 0.30] \\
Household stage & single, couple, family & [0.27, 0.31, 0.42] \\
Ownership stage & first-time, replacement, additional & [0.15, 0.68, 0.17] \\
Primary use case & \begin{tabular}[t]{@{}l@{}}commute, family, luxury,\\performance, mixed\end{tabular} & [0.19, 0.22, 0.14, 0.10, 0.35] \\
\bottomrule
\end{tabular}
\end{table}

\medskip
\subsection{Observable Conditional Approximations}

\begin{table}[H]
\centering
\small
\caption{\(P(\text{income band} \mid \text{age band})\) used in observable-profile sampling.}
\label{tab:income_given_age}
\begin{tabular}{lrrrr}
\toprule
Age band & <60k & 60--100k & 100--180k & 180k+ \\
\midrule
18--25 & 0.42 & 0.36 & 0.17 & 0.05 \\
26--35 & 0.14 & 0.33 & 0.34 & 0.19 \\
36--50 & 0.05 & 0.20 & 0.43 & 0.32 \\
50+ & 0.06 & 0.18 & 0.40 & 0.36 \\
\bottomrule
\end{tabular}
\end{table}

\begin{table}[H]
\centering
\small
\caption{\(P(\text{household stage} \mid \text{age band})\) used in observable-profile sampling.}
\label{tab:household_given_age}
\begin{tabular}{lrrr}
\toprule
Age band & single & couple & family \\
\midrule
18--25 & 0.56 & 0.25 & 0.19 \\
26--35 & 0.32 & 0.31 & 0.37 \\
36--50 & 0.17 & 0.29 & 0.54 \\
50+ & 0.34 & 0.46 & 0.20 \\
\bottomrule
\end{tabular}
\end{table}

\begin{table}[H]
\centering
\small
\caption{\(P(\text{ownership stage} \mid \text{age band})\) used in observable-profile sampling.}
\label{tab:ownership_given_age}
\begin{tabular}{lrrr}
\toprule
Age band & first-time & replacement & additional \\
\midrule
18--25 & 0.52 & 0.42 & 0.06 \\
26--35 & 0.21 & 0.65 & 0.14 \\
36--50 & 0.08 & 0.70 & 0.22 \\
50+ & 0.03 & 0.72 & 0.25 \\
\bottomrule
\end{tabular}
\end{table}

\begin{table}[H]
\centering
\small
\caption{\(P(\text{primary use case} \mid \text{household stage})\) used in observable-profile sampling.}
\label{tab:usecase_given_household}
\begin{tabular}{lrrrrr}
\toprule
Household stage & commute & family & luxury & performance & mixed \\
\midrule
single & 0.30 & 0.08 & 0.17 & 0.10 & 0.35 \\
couple & 0.20 & 0.14 & 0.18 & 0.11 & 0.37 \\
family & 0.12 & 0.43 & 0.08 & 0.06 & 0.31 \\
\bottomrule
\end{tabular}
\end{table}

\medskip
\subsection{Hidden Conditional Mappings}

\begin{table}[H]
\centering
\small
\caption{\(P(\text{decision style} \mid \text{primary use case})\) used in hidden-profile generation.}
\label{tab:decision_given_usecase}
\begin{tabular}{lrrr}
\toprule
Primary use case & analytic & balanced & expressive \\
\midrule
commute & 0.46 & 0.44 & 0.10 \\
family & 0.40 & 0.50 & 0.10 \\
luxury & 0.24 & 0.46 & 0.30 \\
performance & 0.26 & 0.34 & 0.40 \\
mixed & 0.31 & 0.44 & 0.25 \\
\bottomrule
\end{tabular}
\end{table}

\begin{table}[H]
\centering
\small
\caption{\(P(\text{tech-affinity band} \mid \text{age band})\) used in hidden-profile generation.}
\label{tab:tech_given_age}
\begin{tabular}{lrrr}
\toprule
Age band & low & medium & high \\
\midrule
18--25 & 0.10 & 0.36 & 0.54 \\
26--35 & 0.12 & 0.43 & 0.45 \\
36--50 & 0.19 & 0.51 & 0.30 \\
50+ & 0.33 & 0.50 & 0.17 \\
\bottomrule
\end{tabular}
\end{table}

\begingroup
\small
\begin{longtable}{p{0.22\linewidth}p{0.48\linewidth}r}
\caption{\(P(\text{top two priorities} \mid \text{primary use case})\) used in hidden-profile generation.}
\label{tab:priority_given_usecase}\\
\toprule
Primary use case & Priority pair & Probability \\
\midrule
\endfirsthead
\toprule
Primary use case & Priority pair & Probability \\
\midrule
\endhead
\bottomrule
\endfoot
commute & (price, comfort) & 0.30 \\
commute & (comfort, safety) & 0.24 \\
commute & (tech, comfort) & 0.18 \\
commute & (safety, tech) & 0.18 \\
commute & (aesthetics, comfort) & 0.10 \\
family & (comfort, safety) & 0.35 \\
family & (safety, tech) & 0.32 \\
family & (price, comfort) & 0.23 \\
family & (tech, comfort) & 0.10 \\
luxury & (aesthetics, comfort) & 0.34 \\
luxury & (tech, comfort) & 0.30 \\
luxury & (performance, aesthetics) & 0.24 \\
luxury & (comfort, safety) & 0.12 \\
performance & (performance, aesthetics) & 0.50 \\
performance & (tech, comfort) & 0.20 \\
performance & (aesthetics, comfort) & 0.18 \\
performance & (price, comfort) & 0.12 \\
mixed & (price, comfort) & 0.20 \\
mixed & (comfort, safety) & 0.22 \\
mixed & (performance, aesthetics) & 0.13 \\
mixed & (tech, comfort) & 0.17 \\
mixed & (safety, tech) & 0.16 \\
mixed & (aesthetics, comfort) & 0.12 \\
\end{longtable}
\endgroup

\medskip
\subsection{Priority-to-Feature Weight Mapping}

Hidden priority pairs affect valuation through a normalized feature-weight vector over safety, comfort, performance, technology, and aesthetics. The vector starts from a primary-use-case template, shown in \cref{tab:feature_weight_templates}. Analytic decision styles add weight to safety and technology, while expressive decision styles add weight to aesthetics and performance. Each non-price priority in the sampled top-two pair adds weight to its corresponding feature channel; a price priority instead shifts weight toward safety and comfort and away from performance and aesthetics. A small Gaussian perturbation is applied to each channel with a lower floor, and the resulting vector is normalized to sum to one.

\begin{table}[H]
\centering
\small
\caption{Base feature-weight templates used before decision-style and priority adjustments.}
\label{tab:feature_weight_templates}
\begin{tabular}{lrrrrr}
\toprule
Primary use case & Safety & Comfort & Performance & Tech & Aesthetics \\
\midrule
commute & 0.20 & 0.28 & 0.10 & 0.22 & 0.20 \\
family & 0.30 & 0.28 & 0.08 & 0.18 & 0.16 \\
luxury & 0.15 & 0.24 & 0.12 & 0.20 & 0.29 \\
performance & 0.12 & 0.16 & 0.42 & 0.16 & 0.14 \\
mixed & 0.20 & 0.23 & 0.16 & 0.21 & 0.20 \\
\bottomrule
\end{tabular}
\end{table}

For each episode, the selected bundle is also converted into normalized feature-channel signals. Paint, wheels, exterior style, and lighting contribute to aesthetics; upholstery, trim, comfort options, and audio contribute to comfort; technology, safety, and performance options contribute to their matching channels. Option MSRP deltas define the channel mass, with a small structural floor for zero-cost options. The hidden feature-match score is the dot product between the normalized persona feature-weight vector and the normalized bundle feature-channel vector; this score enters the customization value term in the willingness-to-pay calculation.

\medskip
\subsection{Hidden Numeric Mixtures and Linked Rules}

\begin{table}[H]
\centering
\small
\caption{Numeric mixture distributions used in hidden-profile generation.}
\label{tab:hidden_numeric_mixtures}
\begin{tabular}{p{0.27\linewidth}p{0.32\linewidth}p{0.30\linewidth}}
\toprule
Variable & Support & Probability vector \\
\midrule
Price sensitivity & [0.70, 1.00, 1.35] & [0.28, 0.50, 0.22] \\
Aesthetic sensitivity & [0.45, 0.75, 1.05] & [0.24, 0.52, 0.24] \\
Patience & [3, 4, 5, 6] & [0.20, 0.34, 0.30, 0.16] \\
Counter strength & [0.30, 0.55, 0.80] & [0.30, 0.48, 0.22] \\
Walkaway threshold & [0.05, 0.10, 0.18] & [0.42, 0.40, 0.18] \\
Belief obscurity & [0.20, 0.45, 0.70] & [0.30, 0.50, 0.20] \\
Brand loyalty & [0.30, 0.55, 0.80] & [0.24, 0.52, 0.24] \\
Impulsivity & [0.20, 0.45, 0.75] & [0.30, 0.48, 0.22] \\
\bottomrule
\end{tabular}
\end{table}

\begin{table}[H]
\centering
\small
\caption{Income-conditioned reservation-price base distribution.}
\label{tab:reservation_by_income}
\begin{tabular}{lrr}
\toprule
Income band & Mean (USD) & Std (USD) \\
\midrule
<60k & 6,800 & 850 \\
60--100k & 9,200 & 1,100 \\
100--180k & 12,800 & 1,400 \\
180k+ & 17,200 & 1,700 \\
\bottomrule
\end{tabular}
\end{table}

\begin{table}[H]
\centering
\small
\caption{Conditional shift rules applied after initial hidden-variable sampling.}
\label{tab:hidden_conditional_shifts}
\begin{tabular}{p{0.24\linewidth}p{0.20\linewidth}p{0.46\linewidth}}
\toprule
Condition source & Value & Shift applied \\
\midrule
Primary use case & luxury & brand loyalty +0.08 \\
Primary use case & performance & price sensitivity -0.08 \\
Ownership stage & first-time & price sensitivity +0.12; brand loyalty -0.10; walkaway threshold +0.08; patience -1 \\
Ownership stage & replacement & brand loyalty +0.06; walkaway threshold -0.03; patience +1 \\
Ownership stage & additional & price sensitivity -0.05; aesthetic sensitivity +0.08; brand loyalty +0.04 \\
Tech-affinity band & high & brand loyalty +0.04 \\
Priority set contains & price & price sensitivity +0.12 \\
\bottomrule
\end{tabular}
\end{table}

\begin{table}[H]
\centering
\small
\caption{Coupling rules used to keep hidden-variable combinations behaviorally coherent.}
\label{tab:hidden_couplings}
\begin{tabular}{p{0.36\linewidth}p{0.54\linewidth}}
\toprule
Coupling & Rule \\
\midrule
Reservation price from price sensitivity & Multiplicative adjustment with base 1.08, slope -0.18, clipped to [0.75, 1.20]. \\
Walkaway threshold from price sensitivity & Additive adjustment with slope 0.10 around center 1.00. \\
Walkaway threshold from patience & Additive adjustment with slope -0.04 around center 5. \\
Counter strength from belief obscurity & Additive adjustment with slope 0.15 around center 0.50. \\
\bottomrule
\end{tabular}
\end{table}

\clearpage
\section{LLM Prompt Templates}
\label{app:llm_prompt_template}

The following template shows the prompt-v1 contract used for the main LLM experiments. Values inside \texttt{current\_state} are filled from the current observable negotiation state at each seller decision point. The hidden simulator variables described in \cref{sec:simulator_assets} are never inserted into this prompt. The prompt asks for \texttt{move}, \texttt{price\_offer\_usd}, and \texttt{reason}; the evaluation interface requires a non-negative numeric \texttt{price\_offer\_usd} only when \texttt{move} is \texttt{offer}, and retains \texttt{reason} only as trace metadata. The convention of using zero for \texttt{accept} or \texttt{walkaway} is prompt guidance rather than an additional interface requirement.

\begin{Verbatim}[fontsize=\scriptsize,breaklines=true,breakanywhere=true,breaksymbolleft={},breaksymbolright={}]
Return only valid JSON. Do not include Markdown, code fences, or extra text.

{
  "prompt_version": "v1",
  "policy_contract": {
    "role": "You are the seller in a personalized pricing benchmark.",
    "objective": "Maximize expected seller profit from the customization bundle.",
    "profit_formula": "profit_usd = deal_price_usd - estimated_implementation_cost_usd",
    "information_limit": "Use only the observable buyer profile, bundle information,
      and negotiation history below. Hidden willingness-to-pay, hidden preferences,
      patience, and walkaway tendency are not observable.",
    "action_meanings": {
      "offer": "Propose a customization-bundle price in USD. This is not the full vehicle price.",
      "accept": "Accept the buyer's last counter-offer. Use only when last_consumer_offer_usd is not null.",
      "walkaway": "End the negotiation without a deal."
    },
    "accept_precondition": "Use accept only when last_consumer_offer_usd is not null;
      otherwise accept is unavailable in the environment.",
    "decision_guidance": [
      "Prefer profitable deals over no deal.",
      "Do not optimize only for immediate acceptance.",
      "When several rounds remain, use the opportunity to make a profitable but ambitious opening offer.",
      "A first-round offer may be above the expected settlement price if it is still plausible for the bundle.",
      "Avoid offering below estimated_implementation_cost_usd unless strategically necessary.",
      "If the buyer made a counter-offer, compare it with implementation cost and remaining rounds.",
      "If remaining rounds are low, make a realistic final offer or accept a profitable counter.",
      "Treat the observable buyer profile as weak evidence only."
    ]
  },
  "output_contract": {
    "instruction": "Return exactly one JSON object and nothing else.",
    "allowed_actions": ["offer", "accept", "walkaway"],
    "schema": {
      "move": "one of: offer, accept, walkaway",
      "price_offer_usd": "non-negative number required for offer; use 0 for accept or walkaway",
      "reason": "brief string for trace only"
    },
    "example": {
      "move": "offer",
      "price_offer_usd": 5200,
      "reason": "profitable offer adjusted for buyer profile and remaining rounds"
    }
  },
  "current_state": {
    "round": {
      "round_idx": <current seller decision turn>,
      "remaining_rounds": <remaining seller turns after this decision>
    },
    "bundle": {
      "selected_options": [
        {"key": <option key>, "dimension": <dimension>, "msrp_delta_usd": <delta>}
      ],
      "selected_option_keys": [<option keys>],
      "total_msrp_delta_usd": <bundle MSRP delta>,
      "estimated_implementation_cost_usd": <seller-side implementation-cost proxy>,
      "aesthetic_proxy_score": <visible aesthetic proxy>
    },
    "buyer_observable_profile": {
      "age_band": <age band>,
      "income_band": <income band>,
      "household_stage": <household stage>,
      "ownership_stage": <ownership stage>,
      "primary_use_case": <primary use case>
    },
    "negotiation_state": {
      "last_agent_offer_usd": <previous seller offer or null>,
      "last_consumer_response": <latest buyer response>,
      "last_consumer_offer_usd": <buyer counter-offer or null>,
      "history_len": <completed negotiation steps>
    }
  }
}
\end{Verbatim}

Prompt v2 supports the DeepSeek V4 Flash ablation in \cref{tab:supporting_ablations}. It keeps the same output contract and observable \texttt{current\_state} schema as prompt v1, while adding the following fields to make the scenario and state variables more explicit.

\begin{Verbatim}[fontsize=\scriptsize,breaklines=true,breakanywhere=true,breaksymbolleft={},breaksymbolright={}]
{
  "policy_contract": {
    "scenario": "The product is a fixed vehicle customization bundle selected before
      the negotiation. You price only the customization bundle, not the base vehicle.",
    "evaluation_note": "High deal rate alone is not sufficient; low-price immediate
      acceptance can reduce profit.",
    "interaction_dynamics": "After an offer, the buyer may accept, reject, make a
      counter-offer, or walk away. A reject or counter-offer creates a new observation
      in the next round if the episode continues."
  },
  "state_description": {
    "round_idx": "Current seller decision turn, starting from 1.",
    "remaining_rounds": "Number of seller decision turns left after the current one.",
    "selected_options": "Customization options included in the fixed bundle being negotiated.",
    "total_msrp_delta_usd": "Reference retail price increase for the selected customization bundle.",
    "estimated_implementation_cost_usd": "Estimated seller-side cost to provide the selected customization bundle.",
    "aesthetic_proxy_score": "Coarse visible distinctiveness/style proxy for the selected bundle,
      not a hidden preference.",
    "buyer_observable_profile": "Observable buyer demographic and usage-context signals.
      These are weak evidence, not hidden preferences.",
    "last_agent_offer_usd": "Your previous seller offer, or null if no offer has been made.",
    "last_consumer_response": "Buyer response to the previous seller action.",
    "last_consumer_offer_usd": "Buyer counter-offer if one was made; otherwise null.",
    "history_len": "Number of completed negotiation steps so far."
  }
}
\end{Verbatim}

\clearpage
\section{Experiment Uncertainty and Heuristic Details}
\label{app:experiment_uncertainty}

Uncertainty estimates use the episode as the resampling unit. We compute percentile bootstrap intervals with 10,000 resamples and seed 20260511. The intervals in \cref{tab:bootstrap_ci} are computed independently for each method; the table reports method-wise intervals rather than paired intervals for method differences. \Cref{tab:bootstrap_ci} reports 95\% confidence intervals for the two primary metrics used in the main comparison.

\begin{table}[H]
  \centering
  \caption{Episode-level bootstrap 95\% confidence intervals for the main test-set results.}
  \label{tab:bootstrap_ci}
  \small
  \setlength{\tabcolsep}{5pt}
  \begin{tabular}{@{}lrr@{}}
    \toprule
    Method & Deal Rate CI & Avg Profit CI (USD) \\
    \midrule
    Random & [0.5659, 0.5880] & [6,360.82, 6,779.03] \\
    Concession & [0.7171, 0.7368] & [14,554.93, 14,992.80] \\
    DeepSeek V4 Flash & [0.9880, 0.9924] & [6,678.94, 6,820.64] \\
    Kimi K2.6 & [1.0000, 1.0000] & [4,493.44, 4,535.74] \\
    Qwen3.6 Plus & [0.9976, 0.9993] & [5,948.78, 6,010.13] \\
    \bottomrule
  \end{tabular}
\end{table}

The random and concession policies share offer bounds \(L\) and \(U\) computed from the catalog. Let \(m_{\min}\) be the sum of the minimum MSRP delta in each customization dimension, and let \(m_{\max}\) be the corresponding sum of dimension-wise maximum MSRP deltas. The shared bounds are
\begin{equation}
  L = \max(100, 0.4m_{\min}),
  \qquad
  U = \max(L+500, 3.0m_{\max}, 60000).
\end{equation}
The random policy accepts an available buyer counter-offer with probability \(0.12\). It walks away with probability \(0.08\); otherwise it draws an offer uniformly from \([L,U]\).

The concession policy is deterministic up to small Gaussian price noise. Its \(2.20m\) opening anchor and \(1.10m\) floor are set a priori as a transparent aggressive-concession reference, without simulator-based parameter optimization. Let \(m\) denote the selected bundle's total MSRP delta, \(r\) the current seller decision round, and \(R\) the maximum number of seller decision rounds. The round index \(r\) starts at 1, and \(R=r+\texttt{remaining\_rounds}-1\) for the current observation. The policy sets
\begin{equation}
  f = \max(L, 1.10m),
  \qquad
  c = \min\{U, \max(f + 200, 2.20m)\},
\end{equation}
where \(f\) is the floor and \(c\) is the opening anchor. Its round-\(r\) target is
\begin{equation}
  p_r = c + (f-c)\frac{r-1}{\max(1,R-1)} + \epsilon,
  \qquad
  \epsilon \sim \mathcal{N}(0, 100^2).
\end{equation}
When a buyer counter \(b\) is available, the policy accepts it in the final round if \(b \geq f\); otherwise it offers \(\max(b+120, 0.62p_r + 0.38b)\). The final offer is clipped to \([L,U]\). The heuristic returns a floating-point price, and the bargaining protocol rounds submitted offer prices to integer-dollar negotiation issues.

\end{document}